# The competing spin orders and fractional magnetization plateaus of classical Heisenberg model on Shastry-Sutherland lattice: Consequence of long-range interactions

L. Huo[1], W. C. Huang[1], Z. B. Yan[2], X. T. Jia[3], X. S. Gao[1], M. H. Qin[1 a)], and J. -M. Liu[2 b)]

[1]*Institute for Advanced Materials, South China Normal University, Guangzhou 510006, China*

[2]*Laboratory of Solid State Microstructures, Nanjing University, Nanjing 210093, China*

[3]*School of Physics and Chemistry, Henan Polytechnic University, Jiaozuo 454000, China*

**[Abstract]** The competing spin orders and fractional magnetization plateaus of classical Heisenberg model with long-range interactions on a Shastry-Sutherland lattice are investigated using Monte Carlo simulations, in order to understand the fascinating spin ordering sequence observed in $TmB_4$ and other rare-earth tetraborides. The simulation reproduces the experimental 1/2 magnetization plateau at low temperature by considering multifold long range interactions. It is found that more local long range interactions can be satisfied in the 1/2 plateau state than those in the 1/3 plateau state, leading to the stabilization of the extended 1/2 plateau. A mean-field theory on the spin ground states in response to magnetic field is proposed, demonstrating the simulation results. When the energies of the Neel state and the collinear state are degenerated, the former state is more likely to be stabilized due to the competitions among the local collinear spin orders. The present work provides a comprehensive proof of the phase transitions to the Neel state at nonzero temperature, in complimentary to the earlier predictions for the Fe-based superconductors.

---

[a)] Electronic mail: qinmh@scnu.edu.cn
[b)] Electronic mail: liujm@nju.edu.cn

# I. Introduction

During the past decades, frustrated spin systems such as triangular spin-chain system $Ca_3Co_2O_6$ and Shastry-Sutherland (S-S) magnets have attracted widespread interest from both theoretical and experimental approaches due to the fact that fascinating multi-step magnetization behaviors can appear in these systems.[1-6] Up to now, the magnetization plateaus in the former system are generally believed to be caused by non-equilibrium dynamics,[7-9] while those in the latter systems are still under controversy.

The S-S lattice which was first introduced as an interesting example of a frustrated quantum spin system with an exact ground state as early as 1981, can be described as a square lattice with diagonal antiferromagnetic (AFM) coupling $J_1$ in every second square and AFM coupling $J_2$ along the edges of the squares, as depicted in Fig. 1(a).[10] Experimentally, $SrCu_2(BO_3)_2$ with $Cu^{2+}$ ions carrying a quantum spin $S=1/2$ and arranged in a two-dimensional (2d) S-S lattice has triggered an extensive exploration of quantum S-S magnets which exhibit an amazing sequence of magnetization ($M$) plateaus at fractional values of the saturated magnetization ($M_s$).[11] On the other hand, quite a few rare-earth tetraborides $RB_4$ ($R$=Tb, Dy, Ho, Tm, etc.) as another representative of the S-S magnets have been accorded more and more attentions.[5,6,12,13] Similar to $SrCu_2(BO_3)_2$, the complex magnetic structures and their associated physical phenomena in these $RB_4$ compounds under various magnetic field ($h$) at low temperatures ($T$) were identified. For example, the magnetization plateaus at the fractional values of $M_s$ such as $M/M_s$=1/2, 1/7, 1/9, were experimentally observed in $TmB_4$ and several theoretical attempts to understand this interesting phenomenon are available as well.

Unlike $Cu^{2+}$, $Tm^{3+}$ presents a large total magnetic moment ~$6.0\mu_B$, and can be considered as a classical spin system. In addition, $TmB_4$ is of strong easy-axis anisotropy caused by strong crystal field effects. Based on this fact, the magnetization process of the classical AFM Ising model on the S-S lattice was studied using the tensor renormalization-group approach.[14] Under low $h$, either the collinear state [Fig. 2(a)] or the Neel state [Fig. 2(b)] is stabilized depending on the value of $J_1/J_2$. For a certain $T$ range and coupling constants, only a single magnetization plateau at $M/M_s$=1/3 resulting from a particular spin state in which each triangle contains two up-spins and one down-spin [the UUD state, see Fig. 2(c)] was predicted in an intermediate $h$ range. At almost the same time, magnetization pseudoplateaus around $M/M_s$=1/3 were predicted in a classical Heisenberg model on the S-S lattice.[15] Most recently, the ground states of the Ising model on the S-S lattice are investigated and the existence of a single 1/3 plateau has been rigorously proved.[16]

On the other hand, the quantum spin-1/2 Ising-like XXZ model with additional interactions [Fig. 1(b)] on the S-S lattice was visited using the quantum Monte Carlo method, and the experimentally observed magnetization plateau at $M/M_s$=1/2 in the absence of the $M/M_s$=1/3 plateau was reproduced.[17-19] It was indicated that quantum spin fluctuations and long-range interactions may play an important role in the emergence of the $M/M_s$=1/2 plateau, and a ferrimagnetic (FI) state spin arrangement consisting of alternative AFM and ferromagnetic (FM) stripes was recognized [Fig. 2(d)]. In our earlier work, the presence of the $M/M_s$=1/2 plateau was also confirmed when the dipole-dipole interaction is taken into account in the classical Ising model on the S-S lattice.[20,21] Otherwise, a model based on the coexistence of the spin and the electron subsystems was investigated, and the latter subsystem and its interaction with the spin one were believed to be responsible for the plateaus in S-S magnets.[22]

So far, arguments concerning the origin of experimental $M/M_s$=1/2 plateau have not yet reached a consensus. For frustrated spin systems, an effective reduction of the neighboring spin interactions due to the spin frustration may enhance the relative importance of weak interactions. This hints the substantial role of long-range interactions in determining the magnetization behaviors in these frustrated spin systems. For example, a weak FM coupling $J_4$ bond tends to align the connected spin pair parallel with each other, and may stabilize the $M/M_s$=1/2 plateau. Similarly, more local AFM $J_3$ interactions are satisfied in the FI state than those in the UUD state, resulting in the stabilization of the $M/M_s$=1/2 plateau with the increasing $J_3$. In fact, for a quantum spin system, it is confirmed that the FM $J_4$ and AFM $J_3$ couplings are essential to the stabilization of the $M/M_s$=1/2 plateau.[18]

Naturally, one may question whether the same mechanism still holds true for classical spin systems such as $TmB_4$. As a matter of fact, the investigation of the effect of further-neighbor interactions is suggested to eventually explain the plateaus in $TmB_4$ in earlier work.[16] On the other hand, the study of the nontrivial magnetic orders in these systems also plays an essential role in the sense of basic physical research. Most recently, the phase transition from the collinear state to the Neel state at finite *T* was reported in a frustrated AFM model on a square lattice, which is interested in explaining the antiferromagnetic behaviors associated with the Fe-based superconductors.[23] To some extent, this interesting phenomenon probably is observable for S-S magnets. However, as far as we know, few works on this subject have been reported.

In order to clarify this critical issue, we investigate the classical Heisenberg model with the easy-axis anisotropy and the long-range interactions on the S-S lattice. The main $M/M_s=1/2$ plateau can be reproduced when the long-range interactions are included. The phase diagrams obtained by means of the Monte Carlo simulation indicate that both the $J_3$ and $J_4$ interactions have a significant effect on the modulation of the spin configurations. The simulated results at low *T* can be qualitatively interpreted from the spin structures of the ground states for the Ising limit obtained from a mean-field method. In addition, the Neel state is verified to be stable at low temperatures due to the competitions among the local collinear states, which strengthens the conclusion of Wang that the phase transition from the collinear state to the Neel state may occur at finite temperature in the Fe-based superconductors such as P-substituted $LaFeAsO$.[23]

The remainder of this paper is organized as follows: In Sec. II the model and the simulation method will be presented and described. Section III is attributed to the simulation results and discussion of the simulation. The spin structures for the Ising limit at zero temperature will be discussed in Sec. IV, and the conclusion is presented in Sec. V.

**II. Model and method**

The easy-axis anisotropy is ignored in the Heisenberg model, which, on the other hand, is too much emphasized in the Ising limit. The uniaxially anisotropic Heisenberg model seems to be a sound choice for the description of $TmB_4$, as discussed earlier. In the presence of the long-range interactions and *h*, the Hamiltonian can be described as follows:

$$H = J_1 \sum_{diagonal} S_i \cdot S_j + J_2 \sum_{edges} S_i \cdot S_j + J_3 \sum_{\langle i,j \rangle} S_i \cdot S_j$$
$$+ J_4 \sum_{\langle i,j \rangle'} S_i \cdot S_j - h \sum_i S_i^z - D \sum_i (S_i^z)^2 \quad , \quad (1)$$

where the exchange couplings $J_1=1$, $J_2=1/2$, $S_i$ represents the Heisenberg spin with unit length on site $i$, $\langle i,j \rangle$ and $\langle i,j \rangle'$ respectively denote the summations over all pairs on the bonds with $J_3$ and $J_4$ couplings as shown in Fig. 1(b), $h$ is applied along the $+z$ axis, $D=0.4$ is the easy-axis anisotropy constant and $S_i^z$ denotes the $z$ component of $S_i$.

Our simulation is performed on an $L \times L$ ($L=24$) S-S lattice with period boundary conditions using the standard Metropolis algorithm and the parallel tempering algorithm.[24,25] Here, the parallel tempering algorithm is utilized in order to prevent the system from being trapped in the metastable free-energy minima caused by the frustration. We take an exchange sampling after every 10 standard Monte Carlo steps. The simulation is started from the FM state under high $h$, and the $M(h)$ curves are calculated upon $h$ decreasing. Typically, the initial $2\times10^5$ Monte Carlo steps (MCs) are discarded for equilibrium consideration and another $1\times10^5$ MCs are retained for statistic averaging of the simulation.

### III. Simulation results and discussion

In this study, we mainly concern the effect of the long-range interactions on the low-temperature magnetic behaviors of S-S magnets. The calculated $M$ as a function of $J_3$ and $h$ at $T=0.01$ for $J_4=0$ is shown in Fig. 3(a). For small $J_3$ ($J_3<0.1$), $M$ rapidly reaches the first plateau at $M=M_s/3$ resulting from the UUD state when $h$ increases from zero, and then switches to $M_s$ above $h\sim3$. When $J_3$ further increases ($J_3>0.1$), the $M=M_s/3$ step splits into three steps, the $M=0$ step, the $M=M_s/3$ step and the $M=M_s/2$ step. The plateau at $M=0$ is caused by the collinear state, and that at $M=M_s/2$ is caused by the FI state, as reported earlier. The magnetization steps at $M=0$ and $M=M_s/2$ are gradually broadened, while the step at $M=M_s/3$ is narrowed with increasing $J_3$.

Fig. 3(b) shows the simulated phase diagram in the $J_3$-$h$ plane at $T=0.01$, in which the transition points are estimated from the positions of the peaks in the susceptibility $\chi=dM/dh$, following earlier work.[15] At $J_3=0$, the UUD state is stabilized by the magnetic energy when $h$ is applied. The down-spins may flip when $h$ is further increased to the critical field which can be estimated to be $h=3$ for the Ising limit, as verified in our simulation. One may note that the spin pairs on the diagonal $J_3$ bonds tend to be anti-parallel with each other when AFM $J_3$ coupling is taken into account. Compared with the UUD state, the collinear state is stabilized by $J_3$ interaction. Thus, a higher $h$ should be applied to convert the system from the collinear state to the UUD state as $J_3$ increases, leading to the broadening of the magnetization step at $M=0$. To clearly identify the origins of the phase diagram, we respectively calculate the $h$-dependence of the spin-exchange energy $H_1$ from the $J_1$ coupling, $H_2$ from the $J_2$ coupling, $H_3$ from the $J_3$ coupling, the uniaxial anisotropy $H_{an}$, and the Zeeman energy $H_{zee}$ at $T=0.01$ for $J_3=0.3$ [see Fig. 3(c)]. The corresponding magnetization curve is also presented in Fig. 3(d) to help one to understand the results better. The enhancement of the FI state with the increasing $J_3$ can be understood from two parts. On one hand, within certain $h$ range, the energy loss from $H_1$ and $H_2$ due to the phase transition from the UUD state to the FI state is smaller than the energy gain from $H_3$ and $H_{zee}$, leading to the stabilization of the FI state. In addition, the energy gain from $H_3$ due to this transition is increased with the increasing $J_3$, thus the transition shifts toward the small-$h$ side as shown with the red circles in Fig. 3(b), leading to the gradually replacement of the 1/3 plateau by the 1/2 one. On the other hand, the energy loss from $H_3$ due to the phase transition from the FI state to the FM state increases as $J_3$ is increased. So, a larger $h$ will be needed to flip down-spins in the FI state. As a result, when $J_3$ is increased from $J_3=0.1$, the regions of the FI state with the $M=M_s/2$ plateau and the collinear state with the $M=0$ plateau are respectively enlarged, while that of the UUD state with the plateau at $M=M_s/3$ is narrowed.

Similar phenomena can also be observed when an FM $J_4$ coupling is included in this system. Fig. 4(a) shows the calculated $M$ as a function of $J_4$ and $h$ at $T=0.01$ for $J_3=0$. The $M=M_s/3$ step splits into three steps when $J_4$ is increased above -0.05. It is verified that the plateau at $M=0$ is caused by the Neel state which will be discussed in detail in the next section. When $J_4$ is further increased to ~-0.25, the 1/3 step is completely replaced by the $M=0$ step and the $M=M_s/2$ step. In addition, the $M(h)$ curves for larger $J_4$ remain almost the same as that for $J_4=-0.25$. The corresponding phase diagram is shown in Fig. 4(b), and can be clearly understood from the competitions among different energy terms. The energy $H_4$ from the $J_4$ coupling is significantly lost due to the transition from the Neel state to the UUD state as shown in Fig. 4(c), leading to the fact that the transition shifts toward the high-$h$ side with the increasing $J_4$. At the same time, the energy gain from $H_4$ due to the transition from the UUD state to the FI state is increased, resulting in the enhancement of the FI state accompanied by the destabilization of the UUD state. When $J_4$ is increased to -0.25, the UUD state has been completely suppressed, leaving the stabilization of an extended $M=M_s/2$ plateau in the absence of the $M=M_s/3$ plateau. On the other hand, it is shown in Fig. 4(c) and Fig. 4(d) that $H_4$ maintains the same value for the Neel state, the FI state, and the FM state in which all local $J_4$ interactions are satisfied. So, the transition from the Neel state to the FI state and that from the FI state to the FM state respectively occur at stable $h$ irrespective of $J_4$, as clearly shown in Fig. 4(b).

One may note that the perfect collinear state, the Neel state, the UUD state, and the FI state may be partially destroyed near the critical fields for the Heisenberg spin model even at $T$ as low as 0.01, leading to the smoothness of the magnetization curves in our simulation. However, the present work reveals that classical S-S magnets such as TmB$_4$ exhibit complex spin structures which are very sensitive to weak long-range interactions. The experimentally observed magnetization plateau at $M=M_s/2$ can be reproduced when the AFM $J_3$ coupling or/and the FM $J_4$ coupling are taken into consideration. The plateau at $M=0$ stems from either the Neel state or the collinear state, depending on the choice of the exchange interaction coupling constants. To better understand the simulated results, the ground states at zero $T$ for the Ising limit are also discussed based on a mean-field method, as will be found in Sec. IV.

### IV. Magnetic orders at zero temperature: mean-field theory

In fact, the ground-state and the phase boundaries can be qualitatively determined by comparing the Ising energies at different spin configurations which are confirmed from the snapshot of spin configuration in Monte Carlo simulations. Excluding the anisotropy energy (constant here), the energy per site of the Neel state, the collinear state, the UUD state, the FI state, and the FM state can be respectively calculated as follows:

$$E_{\text{Neel}} = J_1/2 - 2 \times J_2 + J_3 + 2 \times J_4, \tag{2}$$

$$E_{\text{col}} = -J_1/2 - J_3 + 2 \times J_4, \tag{3}$$

$$E_{\text{UUD}} = -J_1/6 - 2J_2/3 + J_3/3 + 2J_4/3 - h/3, \tag{4}$$

$$E_{\text{FI}} = 2 \times J_4 - h/2, \tag{5}$$

$$E_{\text{FM}} = J_1/2 + 2 \times J_2 + J_3 + 2 \times J_4 - h. \tag{6}$$

Fig. 5(a) shows the calculated local energies as a function of $h$ for these five states for the Ising limit under $J_3=0$ and $J_4=0$. It is clearly demonstrated that the energies of the UUD state, the FI state and the FM state are degenerated at the saturation field which can be determined to be $h_c=4J_2+J_1=3$ by comparing $E_{\text{UUD}}$ and $E_{\text{FM}}$. As a result, only the $M=M_s/3$ plateau is stabilized in certain $h$ range ($0<h<3$) as shown in Fig. 5(b) in which the corresponding magnetization curves for $D=0.4$ and the Ising limit at $T=0.01$ obtained from Monte Carlo simulation are presented. The local energies for these five states under $J_3=0.1$ and $J_4=-0.1$ are presented in Fig. 5(c). The energy of the collinear state is lower than that of the Neel state when $J_3$ is included, which may be also noted from Eqs. (2) and (3). By comparing $E_{\text{col}}$ and $E_{\text{UUD}}$, one can determine the boundary between the collinear state and the 1/3 plateau state, and the first critical field can be estimated to be $h_{c,1}=4J_3/3-4J_4/3$. The transition shifts toward the high-$h$ side with increasing AFM $J_3$ or/and FM $J_4$. On the other hand, the degeneracy between the UUD state, the FI state and the FM state can be significantly lifted when $J_3$ and $J_4$ are taken into account. When $h$ is further increased, the FI state is stabilized and the second critical field can be estimated to be $h_{c,2}=-2J_3+8J_4+3$ by comparing the energies in Eqs. (4) and (5). $h_{c,2}$ is intensively decreased when the AFM $J_3$ or/and the FM $J_4$ is/are increased. As a result, the main $M=M_s/3$ plateau is gradually replaced by the $M=0$ plateau and the $M=M_s/2$ one, as confirmed in our simulation [Fig. 5(d)]. In addition, the FM $J_4$ coupling is proved to be one of the most efficient ways to stabilize the $M=M_s/2$ plateau, as reported earlier.[18] When $h$ is sufficiently strong, the system should be fully magnetized. Similarly, the upper critical field $h_{c,3}$ is calculated to be $h_{c,3}=2J_3+3$ by comparing $E_{\text{FI}}$ and $E_{\text{FM}}$. So, $h_{c,3}$ is linearly increased with increasing $J_3$, which is irrelevant to $J_4$.

At last, we draw our eyes on the competition between the Neel state and the collinear state under small $h$. It is noted from Eqs. (2) and (3) that these two states are degenerated for zero $J_3$. However, only the perfect Neel state is stabilized for small $h$ at low $T$, as confirmed in our simulation [Fig. 4(b)]. This phenomenon may be qualitatively understood from the competition between two kinds of collinear states with the spin configurations in which the same directed spins are horizontally/vertically arranged. These two states strongly compete with each other in the formation of the perfect collinear state. As a result, the Neel state in which the planar symmetry is still maintained is more likely to be stabilized. This argument is also verified in our simulation by the tracking of the spin configurations at different MCs. The simulation started from an arbitrary state is performed for the Ising limit at $J_3=0$ and $J_4=-0.15$, under zero $h$. The spin configuration at MCs=5000 [Fig. 6(a)] indicates that the local collinear states can be quickly formed. As MCs increase, the competitions among these local collinear states leading to the enlargement of the region with the local Neel state, as shown in Figs. 6(b) and (c). Finally, the system reaches the equilibrium state in which the perfect Neel state dominates, as depicted in Fig. 6(d). Most recently, a possible phase transition from the collinear state to the Neel state is predicted in a 2d square lattice AFM spin model, which is interested in explaining the magnetic behaviors in the Fe-based superconductors.[23] The conclusion may be strengthened by the present work more or less.

## IV. Conclusion

In this study, we have examined the low-temperature magnetic properties of a classical spin model with additional couplings on the Shastry-Sutherland lattice by means of Monte Carlo simulation. The 1/2 magnetization plateau as observed in $TmB_4$ and $ErB_4$ is successfully reproduced when the long-range interactions are taken into account. It is demonstrated that more local long-range interactions are satisfied in the 1/2 plateau state than those in the 1/3 plateau one, leading to the stabilization of the extended 1/2 plateau. The origins of these interesting magnetic orders are discussed in details, and are confirmed by the ground state analysis based on a mean-field method. It is indicated that even weak long-range interactions may have a significant effect on the step-like magnetization feature for the classical S-S magnets. In addition, the competitions between the Neel state and the collinear state is discussed, and the former one is confirmed to be stabilized when the energies of these two states are degenerated. This simulated result provides evidence to the conclusion of the earlier work in which a phase transition to the Neel state is predicted in the Fe-based superconductors such as P-substituted LaFeAsO.

**Acknowledgements**:

This work was supported by the Natural Science Foundation of China (11204091, 11274094, 50832002), the National Key Projects for Basic Research of China (2011CB922101), China Postdoctoral Science Foundation funded project (2012T50684, 20100480768), and the Priority Academic Program Development of Jiangsu Higher Education Institutions, China.

**FIGURE CAPTIONS**

Fig.1. (color online) Effective models on the Shastry-Sutherland lattice with (a) the diagonal coupling of $J_1$, and $J_2$ along the edges of the squares, (b) the additional interactions $J_3$ and $J_4$.

Fig.2. (color online) Spin configurations in (a) the collinear state, (b) the Neel state, (c) the UUD state, and (d) the FI state. Solid and empty circles represent the up-spins and the down-spins, respectively.

Fig.3. (color online) (a) Magnetization $M$ versus $J_3$ and magnetic field $h$. The parameters are $L=24$, $T=0.01$ and $J_4=0$. (b) Phase diagram of the magnetization plateau in the $h$-$J_3$ plane. The calculated (c) $H_1$, $H_2$, $H_3$, $H_{an}$, $H_{zee}$ and (d) magnetization $M/M_s$ as a function of $h$ at $T=0.01$ for $J_3=0.3$.

Fig.4. (color online) (a) Magnetization $M$ versus $J_4$ and magnetic field $h$. The parameters are $L=24$, $T=0.01$ and $J_3=0$. (b) Phase diagram of the magnetization plateau in the $h$-$J_4$ plane. The calculated (c) $H_1$, $H_2$, $H_4$, $H_{an}$, $H_{zee}$ and (d) magnetization $M/M_s$ as a function of $h$ at $T=0.01$ for $J_4=-0.15$.

Fig.5. (color online) The local energies as a function of $h$ for the Ising limit for (a) $J_3=0$ and $J_4=0$ and (c) $J_3=0.1$ and $J_4=-0.1$. (b) and (d) are the correspondingly simulated magnetization curves at $T=0.01$.

Fig.6. (color online) Typical MC snapshot of the spin structure at $T=0.01$ for the Ising limit at (a) MCs=5000, (b) MCs=40000, (c) MCs=80000, and (d) MCs=120000. The other parameters are $J_3=0$, $J_4=-0.15$ and $h=0$. Solid and empty circles represent the up-spins and the down-spins, respectively. The blue lines are guides to the eyes.

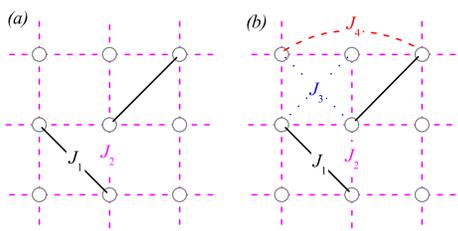

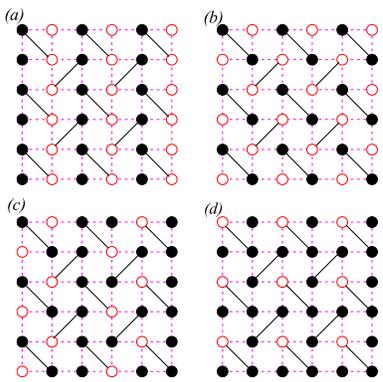

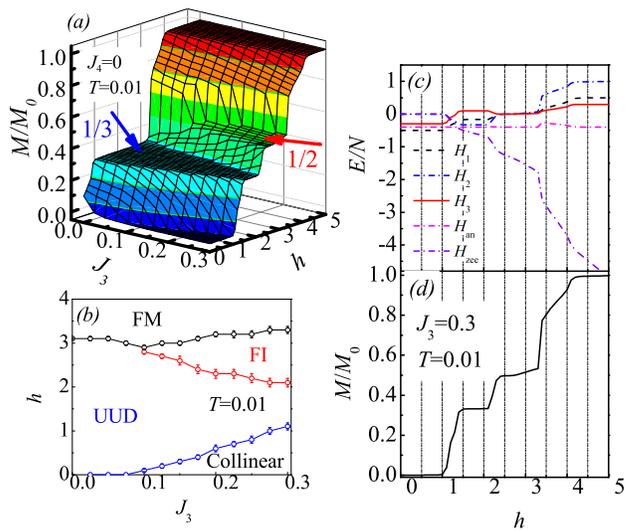

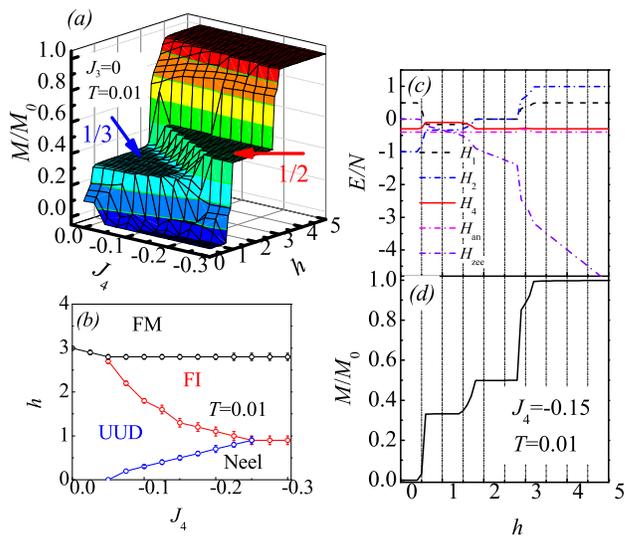

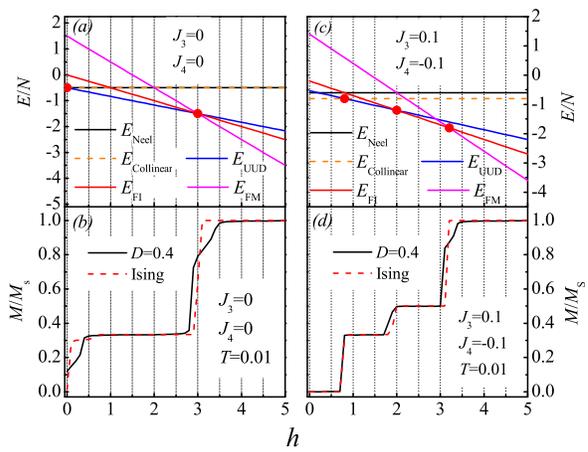

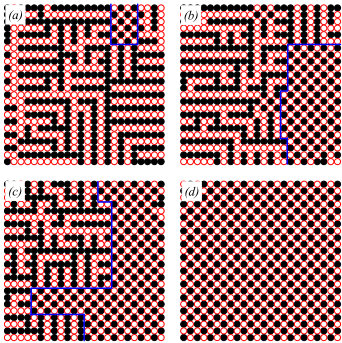